\documentclass[fleqn,10pt]{wlscirep}

\usepackage[utf8]{inputenc}
\usepackage[T1]{fontenc}
\usepackage{hyperref}       
\usepackage{url}            
\usepackage{booktabs}       

\title{Simulated redistricting plans for the analysis and evaluation of redistricting in the United States}

\author[1]{Cory McCartan}
\author[2]{Christopher T. Kenny}
\author[2]{Tyler Simko}
\author[3]{George Garcia III}
\author[4]{Kevin Wang}
\author[4]{Melissa Wu}
\author[5]{Shiro Kuriwaki}
\author[1,2,*]{Kosuke Imai}

\affil[1]{Department of Statistics, Harvard University}
\affil[2]{Department of Government, Harvard University}
\affil[3]{Department of Economics, Massachusetts Institute of Technology}
\affil[4]{Harvard College}
\affil[5]{Department of Political Science, Yale University}
\affil[*]{\texttt{imai@harvard.edu}}

\usepackage{color}
\usepackage{fancyvrb}

\DefineVerbatimEnvironment{Highlighting}{Verbatim}{commandchars=\\\{\}}
\usepackage{framed}
\definecolor{shadecolor}{RGB}{248,248,248}
\newenvironment{Shaded}{\begin{snugshade}}{\end{snugshade}}

\newcommand{\AttributeTok}[1]{\textcolor[rgb]{0.77,0.63,0.00}{#1}}

\newcommand{\ConstantTok}[1]{\textcolor[rgb]{0.00,0.00,0.00}{#1}}

\newcommand{\FunctionTok}[1]{\textcolor[rgb]{0.00,0.00,0.00}{#1}}

\newcommand{\NormalTok}[1]{#1}

\newcommand{\OtherTok}[1]{\textcolor[rgb]{0.56,0.35,0.01}{#1}}

\newcommand{\SpecialCharTok}[1]{\textcolor[rgb]{0.00,0.00,0.00}{#1}}

\newcommand{\StringTok}[1]{\textcolor[rgb]{0.31,0.60,0.02}{#1}}

\newcommand{\fiftydata}{\textsc{50stateSimulations}}
\usepackage{booktabs}
\usepackage{makecell}
\usepackage{longtable}
\definecolor{Periwinkle}{HTML}{7977B8}

\begin{abstract}
This article introduces the \fiftydata, a collection of simulated
congressional districting plans and underlying code
developed by the Algorithm-Assisted Redistricting Methodology
(ALARM) Project. The \fiftydata~allow for the evaluation of enacted and
other congressional redistricting plans in the United States.
While the use of redistricting simulation algorithms has
become standard in academic research and court cases, any simulation
analysis requires non-trivial efforts to combine multiple data sets,
identify state-specific redistricting criteria, implement complex simulation
algorithms, and summarize and visualize simulation outputs. We have
developed a complete workflow that facilitates this entire process of
simulation-based redistricting analysis for the congressional districts of
all 50 states. The resulting \fiftydata~include ensembles of simulated
2020 congressional redistricting plans and necessary replication data.
We also provide the underlying code, which serves as
a template for customized analyses. All data and code are free and publicly
available. This article details the design, creation, and validation of the data.
\end{abstract}

\begin{document}

\flushbottom
\maketitle

\def\tightlist{}

\keywords{
    keyword 1
   \and
    keyword 2
  }

\thispagestyle{empty}

\renewcommand*{\thefootnote}{\fnsymbol{footnote}}

\hypertarget{background-summary}{%
\section*{Background \& Summary}\label{background-summary}}
\addcontentsline{toc}{section}{Background \& Summary}

Redistricting---the process of redrawing electoral district boundaries following the constitutionally mandated decennial census---has substantial impacts on representation, voting rights, and governance in the American political system.
As a fundamentally political process, redistricting has also been manipulated to fulfill partisan ends.
The detection of gerrymandering, i.e., the intentional redrawing of district boundaries to unduly advantage or disadvantage a certain group of voters, is of immense importance among scholars, policymakers, federal and state courts, and the general public.
Just within the 2020 redistricting cycle, at least 72 cases across 26 states have been filed to challenge congressional and state legislative redistricting plans as racially discriminatory and/or gerrymandered for partisan gain \cite{Brennan:2022}.

Evaluating a redistricting plan, however, is not a straightforward task.
It requires the analyst to take into account federal requirements as well as each state's redistricting criteria and particular political geography.
Comparing the partisan bias of a plan for Texas with that of a plan for New York, for example, is likely misleading given numerous differences in redistricting requirements, demographics, geography, candidates, and other state-specific factors.
Comparing a state's current plan to its past plans to detect gerrymandering is also problematic because its demographics, politics, and institutions may have changed over the intervening time period between plans.
Often, the implicit goal in these cross-state and cross-time comparisons is to assess how the redistricting plan compares to other possible plans in the same state.
Under this approach, for example, a redistricting plan can be considered as a partisan gerrymander if it constitutes an outlier, relative to a sample of alternative plans that satisfy the same set of statutory guidelines and requirements, with respect to certain partisan bias metrics.\cite{KatzEtAl:2020}

Simulation algorithms have been designed to generate these alternative redistricting plans.
Recent advances in computing and methodologies, along with the increasing availability of granular data about voters and elections, have led to the development of redistricting simulation methods that can incorporate federal laws and guidelines, state-specific rules, and election data \cite{carter2019,deford2019,autry2020,fifield2020mcmc,McCartanEAl:2021}.
These simulation methods have gained widespread use in federal and court cases challenging redistricting plans.
During this redistricting cycle alone, evidence based on simulation algorithms has been used in courts across a large number of states including Alabama, Georgia, New York, North Carolina, and Ohio.
Indeed, simulation analyses have become a standard method for evaluating redistricting plans in academic and judicial settings.

For most analysts, however, performing a redistricting simulation analysis is a complex and laborious task.
To begin, the analyst first must put together a detailed geographic dataset combining boundary geometries, legislative district plans, demographic information from the Census, and election data.
These data generally come from different sources, and may not naturally overlap with each other.
Once the necessary data are tidied and joined, the analyst must then identify the redistricting criteria that should constrain the alternative plans.
These criteria are typically based on federal and state laws and need to be formalized as statistical constraints for redistricting simulation algorithms.
Next, the analyst must implement a redistricting simulation algorithm to generate a representative sample of alternative redistricting plans that are both diverse and conform to the redistricting criteria.
Lastly, they must compare the sampled plans on a wide-ranging set of metrics that past researchers have developed.

We created the \fiftydata to aid scholars, policymakers, data journalists, and citizen data scientists alike in performing redistricting simulation analyses.
The \fiftydata offer a set of data and computing tools that drastically cut down the complexity and time required to conduct such analyses.
The \fiftydata include tidied 2020 decennial Census joined with retabulated election data from the Voting and Election Science Team (VEST), a representative sample of simulated 2022 Congressional redistricting plans for all 50 states, and a suite of software packages to visualize, explore, and simulate redistricting plans \cite{Census:2022,VEST:2022}.
Any analyst can use these alternative redistricting plans immediately to evaluate the potential bias of an enacted plan or any other counterproposal.
In addition, the \fiftydata include the code used to generate these simulated plans, which can serve as templates for custom redistricting simulation analyses.
Everything in the \fiftydata is open-source and reproducible.
In this paper, we provide an overview of the workflow for building the \fiftydata, describe its contents, and illustrate its use.

\hypertarget{methods}{%
\section*{Methods}\label{methods}}
\addcontentsline{toc}{section}{Methods}

When states enact Congressional district plans, they make a series of discretionary decisions.
How many counties and towns should be split?
How compact should districts be?
A major benefit of using simulations is the ability to incorporate such redistricting criteria in a transparent fashion.
We illustrate the simulation process that generates the \fiftydata by using Georgia's Congressional redistricting plan as an example.
An overarching goal in this process is to generate a representative set of alternative plans that conforms to the redistricitng criteria of that state.

After the 2020 decennial Census, the state of Georgia was allocated 14 congressional districts, each of which elects a single member of the US House of Representatives by plurality vote in 2022 and subsequent general elections.
The Georgia State Legislature has the authority to decide how these particular districts are drawn, and the plan is enacted after the Governor's signature.
How exactly these districts are drawn is politically consequential, but neither the state constitution of Georgia nor the Georgia Code specifies legal requirements for Congressional redistricting.
The map-drawing process, however, adheres in principle to the guidelines established by the state legislature, the authority responsible for redistricting in Georgia.
Under the 2021-22 guidelines for Georgia's House Legislative and Congressional Reapportionment Committee\cite{georgialegcmte}, districts must: (1) be contiguous, (2) have equal populations, (3) be geographically compact, (4) preserve county and municipality boundaries as much as possible, and (5) avoid the unnecessary pairing of incumbents.
Our simulations account for all of these criteria with the exception of incumbency pairing. The criteria used in Georgia are fairly standard, while other states like Colorado (which requires the creation of competitive districts) and Ohio (which has specific requirements about which counties and municipalities can be split and how often) require much more specific criteria that we incorporate into our simulations. Supplementary Table 1 shows the legal sources and redistricting criteria we included for every state.

Our simulations in Georgia also include a constraint based on the 1965 Voting Rights Act (VRA).
According to the VRA, minority racial groups that are polarized from the majority should be arranged in districts that can elect their members of choice.
In practice, this means that districts should have sufficient numbers of minority constituents as measured, for example, by the proportion of the Voting Age Population that is Black (BVAP).
In Georgia, we incorporate a consideration of the VRA by penalizing districts that have a BVAP lower than 52 percent.
We describe the reasoning behind this choice in the Technical Validation section, and list specifications for other states in Supplementary Table 1.
This sets a soft, probabilistic target for the minority proportion in a district based on the enacted plan, but allows the proportion to vary below or above the value set.
We then simulate alternative redistricting plans under these criteria.

The first step of the redistricting simulation workflow is to assemble precinct-level shapefiles with associated demographic data.
The \fiftydata contains the ALARM Project's 2020 Redistricting Data Files that consist of the tidied 2020 decennial Census and statewide election data from the Voting and Election Science Team (VEST)\cite{VEST:2022}.
The VEST data are widely used in academic research\cite{allcott2020polarization, grossman2020political}, litigation (e.g., ``An Evaluation of the Partisan Fairness of Ohio's February 24, 2022 State Legislative Districting Plan'', report of Christopher Warshaw in \emph{League of Women Voters of Ohio v. Ohio Redistricting Commission} (2020)), and public projects (such as in Dave's Redistricting App or PlanScore).
The election data are tidied by estimating VEST's collection of precinct shapes to their underlying 2010 blocks using \cite{geomander}, then crosswalked to 2020 blocks \cite{VESTcw, PL94171} and aggregated to voting districts.
Election data are first matched to 2010 blocks, as precincts are largely made of census blocks, which are mostly stable for the decade \cite{amos:mcdo:watk:17}.
We also include data about municipality boundaries, which are obtained from Census block files.
We acquire the shapefiles for Georgia's enacted congressional plan from the General Assembly website (\url{https://www.legis.ga.gov/joint-office/reapportionment}) and join them to the rest of the data.

We incorporate the redistricting criteria set by the legislative committee into our simulation analysis using both hard constraints, which ensure that all simulated plans meet the specified criteria, and soft constraints, which encourage simulated plans to meet the specified criteria without enforcing hard limits.
We use a Sequential Monte Carlo (SMC) redistricting algorithm \cite{McCartanEAl:2021} to obtain a representative sample of alternative redistricting plans under these redistricting criteria. The SMC algorithm enforces two hard constraints: contiguity and limited deviation from population parity, which reflect universal redistricting criteria.
The algorithm by default also includes a soft constraint which encourages district compactness, according to a particular graph-theoretic measure of compactness.\footnote{Specifically, the algorithm samples plans according to the number of spanning forests which can be drawn on the district adjacency graph. This graph-theoretic quantity correlates strongly with the total perimeter of the districts, and thus also with non-graph-based compactness measures which are based on perimeter length, such as Polsby--Popper.}
This compactness constraint can be tuned slightly upwards or downwards, which we do in states with specific compactness requirements (see the Appendix for details).
The defaults, however, produce districts with a range of compactness values that generally span the range of historical values for congressional districts, and so in most states we do not tune the constraint.
For the population equality constraint, we set the maximum deviation from the population parity to be 0.5\%, which corresponds to approximately 3,826 people in Georgia.
Although enacted redistricting plans often achieve exact population parity across districts, the use of the 0.5\% threshold is appropriate because our simulation analysis is based on precincts (the smallest units for which the electoral data are available) rather than the more granular Census blocks.

Additionally, we add a hard constraint to the algorithm to limit the number of counties and municipalities that are split by districts.
This hard constraint is incorporated as part of the SMC algorithm itself, and limits the number of splits of provided administrative units to one less than the number of districts.
It is up to the analyst, however, to decide what administrative units to provide to the algorithm.
Many states, including Georgia, value the preservation of both county and municipality boundaries, and we operationalize this constraint in the following way.
We start by providing county boundaries to the SMC algorithm; this will limit the number of county splits.
However, counties with populations larger than a congressional district must necessarily be split.
In these counties (which are Cobb, Fulton, and Gwinnett counties, in Georgia), we use municipalities (specifically, Census Designated Place boundaries) as the administrative units.
Across the whole state, then, the SMC algorithm will limit the number of county (in lower-population areas) and municipality (in large counties) splits to one less than the number of districts.
We check the sampled plans against the enacted plan to ensure that the sampled plans perform as well or better, on average, than the enacted plan.
In some other states, there are specific rules about the number of counties and municipalities that may be split, and we ensure every sampled plan meets those rules, either by redifining the administrative boundaries, adding further soft constraints to encourage fewer splits, or by subsetting the sampled plans to those which satisfy the requirements.

Lastly, we use a soft constraint to encourage sampling redistricting plans that have the same number of Black opportunity districts as the enacted plan.
In other states, this constraint may be applied separately for Black and Hispanic voters, or collectively applied to minority voters overall if no one racial group has sufficient numbers to form a majority in a district.
The strength of this constraint must be carefully tuned to ensure the algorithm functions correctly while producing the specified range of majority-minority or opportunity districts.
The Technical Validation contains further details on how we consider the nuances of the VRA.

We simulate 20,000 alternative redistricting plans for Georgia, ensuring that the sample has converged into a stable distribution by comparing two parallel runs of the SMC algorithm.
We then thin this set of plans to 5,000 plans for our final product.

In most states, we obtain two independent samples of 2,500 simulations in parallel and combine these samples to generate a sample of 5,000 simulated plans.
No thinning is necessary in these states.
In some large states such as Georgia that have more geographical complexity, we sample more than 5,000 plans in total so that the simulation algorithm converges, according to numerical diagnostics (see Technical Validation section).
Once these diagnostics suggest convergence, we randomly sample 5,000 plans from the original set of simulated plans.
This is done to maintain consistency across states and limit the memory usage for simulated data, eliminating a potential computational burden for users.
A redistricting plan is essentially an assignment of each precinct in the state to a district number.
Each alternative plan offers a distinct set of assignments (with some possible duplication), so that each simulated district covers a different geographic area and is therefore associated with different demographic and partisan characteristics.
Finally, we post-process the plans so that their numbering matches that of the enacted plan---district 1 in any simulated plan will roughly correspond to district 1 in the actual plan, and so on.

\begin{figure}
\includegraphics[width=1\linewidth]{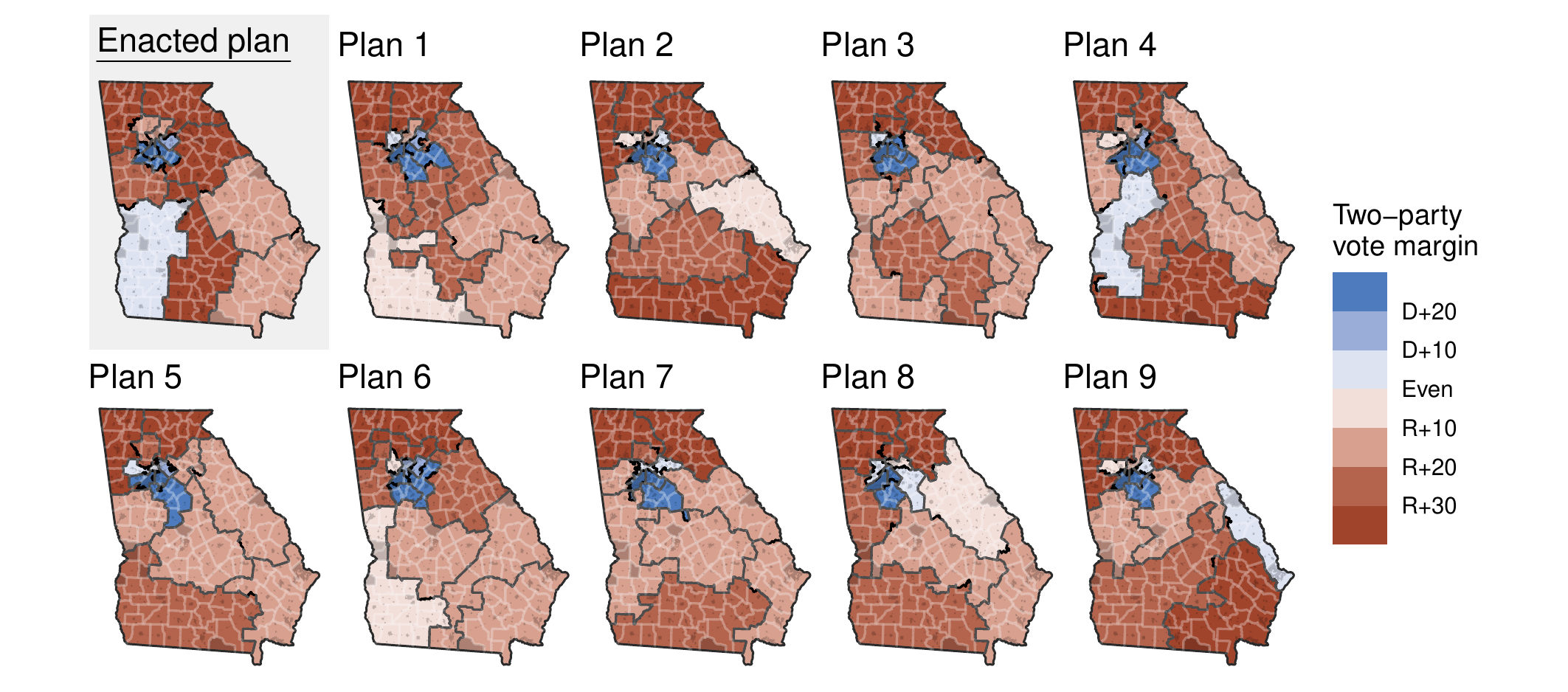} \caption{The enacted Georgia congressional plan (top left) and 9 out of the 5,000 alternative plans we provide in this dataset. Gray areas indicate Census-designated places (including cities and towns), white lines indicate counties, and dark gray lines indicate district assignment. Each district is colored by its partisan lean computed as an average of historical statewide election results (2016--2020) within the sampled district, where D+20 indicates the average Democratic candidate wins by 20 percentage points or more over the average Republican candidate. Because each alternative plan can assign different districts to each precinct, the district-level two-party vote can also differ.}\label{fig:ga-samp}
\end{figure}

As an example, we show 9 out of the 5,000 alternative plans of Georgia in Figure \ref{fig:ga-samp}.
The district assignment that the Georgia state legislature finalized is also shown for comparison.
We then characterize each simulated district by its demographics and partisan lean according to historical election data.

The final dataset is a series of alternative district assignments for each precinct in the state, along with the demographic and political characteristics of those alternative plans.
We show how to extract and use these variables in the next section.

\hypertarget{data-records}{%
\section*{Data Records}\label{data-records}}
\addcontentsline{toc}{section}{Data Records}

The \fiftydata can be found in the Harvard Dataverse\cite{50statesDataverse} at \url{https://doi.org/10.7910/DVN/SLCD3E}.
Each state has four files that follow the \texttt{\{state\}\_cd\_\{yyyy\}\_\{type\}} naming convention, with \texttt{state} indicating the state abbreviation, \texttt{cd} indicating the congressional districts, \texttt{yyyy} indicating the 4-digit year of the redistricting cycle, and \texttt{type} taking on (1) \texttt{doc} for the markdown documentation, (2) \texttt{map} for the \texttt{redist\_map} object, (3) \texttt{plans} for the \texttt{redist\_plans} object, and (4) \texttt{stats} for the summary statistics.
These files are organized into folders by state in the \emph{Tree View} option on the Dataverse website.

Documentation for each state's redistricting simulation methodology is outlined in the \texttt{\_doc.html} file.
We record information about the state's legal redistricting requirements and a description of the algorithmic constraints that we implemented to comply with those requirements.
We list the sources from which we obtain our geographical, population, election, and enacted plan data.
We also describe any pre-processing notes, the number of plans simulated, and any special techniques needed to produce the sample.
The documentation includes a brief description of each file in the state's folder and explanations of each summary statistic in a listed format.

Underlying each simulation is a \texttt{redist\_map} object (a custom object defined in the \emph{redist} package \cite{redist}), which has pre-merged geographic, demographic, and electoral data at each voting tabulation district (or precinct).
It is a shapefile that contains the geographic coordinates and adjacency of each unit.
The \texttt{redist\_map} object contains geographical data such as \texttt{GEOID} (precinct/block unique identifier), \texttt{adj} (an adjacency graph that records which precincts are geographically adjacent to one another), \texttt{state} (state abbreviation), \texttt{county} (county name), \texttt{muni} (municipality identifier), \texttt{county\_muni} (concatenation of \texttt{county} and \texttt{muni}), \texttt{cd\_2010} (congressional district number assignment in the 2010 enacted plan), \texttt{cd\_2020} (congressional district number assignment in the 2020 enacted plan), and \texttt{vtd} (voting district identifier).

The \texttt{redist\_map} object also contains Census and election statistics measured at the precinct-level that will later be aggregated up to districts.

\begin{itemize}
\tightlist
\item
  Population and demographic information comes from the Census PL94-174 file: with \texttt{pop} indicating the entire population and \texttt{vap} indicating voting-age population. The size of demographic subgroups are also included for targeting effective minority districts in some states. Racial subgroups are denoted by \texttt{\_hisp} (Hispanic or Latino of any race), \texttt{\_white} (White alone, not Hispanic or Latino), \texttt{\_black} (Black or African American alone, not Hispanic or Latino), \texttt{\_aian} (American Indian and Alaska Native alone, not Hispanic or Latino), \texttt{\_asian} (Asian alone, not Hispanic or Latino), \texttt{\_nhpi} (Native Hawaiian and Other Pacific Islander alone, not Hispanic or Latino), \texttt{\_other} (some Other Race alone, not Hispanic or Latino), and \texttt{\_two} (population of two or more races, not Hispanic or Latino).
\item
  Electoral information relies on the variables from VEST. Statewide offices and their election schedules vary by state. In Georgia, for example, VEST includes elections for President (2016, 2020), US Senate (2016, 2020), Governor (2018), Attorney General (2018), and Secretary of State (2018). The data from VEST uses the naming convention \texttt{\{office\}\_\{year\}\_\{party\}\_\{candidate\}} to measure the vote totals of each candidate at the precinct-level: \texttt{office} indicates the office abbreviation, with options including President (\texttt{pre}), United States Senate (\texttt{uss}), Governor (\texttt{gov}), Attorney General (\texttt{atg}), Secretary of State (\texttt{sos}); \texttt{year} indicates the last two digits of the election year; \texttt{party} indicates either the Republican candidate (\texttt{rep}) or Democratic candidate (\texttt{dem}); \texttt{candidate} indicates the first three letters of the candidate's last name.
\item
  We then summarize these VEST variables as \texttt{arv\_\{year\}} (average vote counts for Republican candidates in that year), \texttt{adv\_\{year\}} (average vote counts for Democratic candidates), \texttt{nrv} (average vote counts for Republican candidates across all available election years), and \texttt{ndv} (average vote counts for Democratic candidates across all available election years).
\end{itemize}

We save the \texttt{redist\_map} object as a compressed RDS file, a native R format that retains the basic attributes of the redistricting problem, such as the number of districts to draw and the population parity constraint they should satisfy.
It also retains the shapefile and adjacency information for each precinct.

The set of plans simulated from this map object are stored in a \texttt{redist\_plans} object (another custom object defined in the \emph{redist} package).
Each state's simulations are constructed differently, following the specific rules that govern redistricting in each state (see the Technical Validation for details).
Each \texttt{redist\_plans} object contains the assignment of districts to each unit of geography (in our case the precinct).
Each alternative plan and a reference plan are encoded by \texttt{draw} and the district is indexed by \texttt{district}.
For a redistricting problem of 1,000 precincts to be assigned to 5 districts, if we simulate 5,000 alternative plans and compare it with the plan enacted by the legislature, the \texttt{redist\_plans} object contains \((5000 + 1) \times 5 = 25,005\) rows.
Simulations are conducted over multiple independent runs of the redistricting algorithm, which is useful for diagnostic purposes, and these runs are identified by the \texttt{chain} column.

Our simulation output is most commonly used to compare a district-level or plan-level summary statistic of a particular plan to the entire distribution of that summary statistic from the simulated plans.
We provide a plain-text comma-separated values (CSV) file that contains these summary statistics for each of the 5000 plans or simulated districts.
This file does not include any R-specific attributes, and can be used in any programming or spreadsheet software.
If users have a statistic for a proposal that is not in our data, e.g., the Efficiency Gap metric for a remedial map proposed in court, they can easily compare how that number compares to our reference distribution by loading the summary statistics in a spreadsheet or data analysis program of their choice.

The summary statistics file includes the totals of any variables included in the \texttt{redist\_map} population and election data (such as the total population of racial subgroups in each district).
The difference is that \texttt{redist\_map} provides precinct-level measures of these data, while the \texttt{redist\_plans} object and summary file show the district-level totals of these building blocks as they are aggregated into different district arrangements.
It also contains plan-level statistics for traditional redistricting criteria such as: the maximum population deviation among plan districts (\texttt{plan\_dev}), plan-level compactness according to the fraction of edges kept (\texttt{comp\_edge}), compactness according to the Polsby-Popper score (\texttt{comp\_polsby}).\cite{polsby1991third}

While many different compactness measures are used today, we have adopted these two as baseline summary statistics, in part because they are widely used in academic work and litigation and are computationally efficient to calculate.
The Polsby-Popper score is perhaps the most widely-used compactness measure.
The fraction of edges kept is a graph-theoretic measure and thus, unlike the Polsby-Popper score, it is invariant to changes in the resolution of the shapefile or the inclusion or exclusion of certain water areas from precinct boundaries.
This compactness measure is also closely related to the graph-theoretic measure that is used in the SMC algorithm's built-in compactness constraint.
We note that among the simulated plans, there is a high degree of correlation between both of these compactness measures.
For users who require additional compactness information, the \textit{redist} and \textit{redistmetrics} packages provide functionality to easily compute many of these additional measures for the simulated plans.

Of interest for detecting partisan gerrymandering are normal Democratic share (\texttt{ndshare}), average Democratic vote share (\texttt{e\_dvs}), probability that a seat is represented by a Democrat (\texttt{pr\_dem}), expected number of Democratic seats for each plan (\texttt{e\_dem}).
We estimate the expected number of Democratic seats by first determining the winner of election within each simulated district using the precinct-level vote shares for a particular historical statewide race.
We then compute the number of districts Democratic party candidates are expected to win.
Taking the average of this number across all statewide races gives the expected number of Democratic seats.
It is important to note that because a different set of statewide elections are available in each state, partisan summary statistics may not be directly comparable across states.
We encourage users to consider modeling election results using a baseline partisanship measure derived from presidential elections, which have results available for all states.

These summary statistics of elections can be further transformed to compute common measures of partisan gerrymandering: The difference from the expected number of seats won by a party; the deviation from partisan symmetry,\cite{KatzEtAl:2020} which estimates the difference in each party's seat share if they each won 50\% of the statewide vote (the variable \texttt{pbias}); and the efficiency gap,\cite{stephanopoulos2015partisan} which counts the difference in the number of wasted votes for each party averaged across all available elections (the variable \texttt{egap}).
For partisan bias measures, a positive value implies a pro-Republican bias while a negative value indicates a pro-Democratic bias.

\hypertarget{technical-validation}{%
\section*{Technical Validation}\label{technical-validation}}
\addcontentsline{toc}{section}{Technical Validation}

In preparing each state's simulation, we studied the laws and regulations of each state's redistricting process and operationalized them through simulation constraints in the \texttt{redist} package.
These goals for compliance must be balanced with maintaining a diverse sample of plans.

To ensure that the simulated plans are drawn from a population of valid alternatives, we check the population deviation at the voting district level, two measures of compactness (the fraction of edges kept compactness and Polsby-Popper compactness), number of county splits, number of municipality splits, and the minority voting age population in each district.
We assess compliance with such traditional redistricting criteria by asking if the distribution of statistics of the alternative plans are in line with the enacted plan or historical plan.

The process is iterative. If initial simulations appear overly non-compact or appear to split administrative boundaries excessively, we consider strengthening the soft constraints for these metrics.
However, we do not strictly tune our constraints to the precise compactness of the enacted map.
As mapdrawers know that compactness will be used to evaluate redistricting maps, compactness can be used to disguise a gerrymander.
For example, in Florida and North Carolina, our final simulations are less compact than the enacted map.
In contrast, in states such as Illinois, our final simulations are more compact, as compactness was subjugated to develop a Democratic advantage in the enacted plan.

Statistics about the size of the racial minority population in each district is fraught with minority vote dilution and racial gerrymandering litigation stemming from the implementation of the Voting Rights Act of 1965 (VRA).
We ensure that the simulated plans would be \emph{comparably} compliant with Section 2 of the VRA, which stipulates that minority voters cannot be deprived of the ability to be represented by their district representative of their choice.\cite{grofmaneffective}
While compliance is a legal question, we use this to guide the simulations under the assumption that the enacted plan complies with the VRA.
In states with a history of litigation or preclearance, we therefore ensure that simulated plans have equal opportunity to elect candidates of choice as in the enacted plan by checking whether districts with large minority populations end up with a higher vote share for the party that the minority population is known to support.
For example, in Georgia, external estimates suggest that Black voters overwhelmingly support the Democratic party, so we verify that districts with relatively more Black voters also elect Democratic candidates according to the historical election data.

Simulating plans which are VRA compliant can be a difficult procedure, as one must determine which districts are minority opportunity districts.
There is neither academic nor legal consensus on best practices.
Chen and Stephanopoulos (2021)\cite{chen:step:21} considers a three step test: (1) the minority-preferred candidate must win the general election, (2) there must be more minority voters for the winner than white voters, and (3) minority voters count towards criteria (2) only if the groups prefer the same candidate.
Becker et al.~(2021)\cite{beck:etal:21} considers a different approach, defining a model of district performance on the logit scale.
Our approach is closer to that of Chen and Stephanopoulos (2021)\cite{chen:step:21}.
More precisely, we define a minority opportunity district as a district where (1) the minority voting age population proportion is in ranges which produce candidates of choice from minority voters (typically Democrats) and (2) minority voters are expected to contribute more to the winning coalition than white voters.
This means that we do not target a particular racial lower bound while simulating districts.
After plans are sampled, we check that all plans would have at least as many minority opportunity districts as the enacted plan, again ensuring that the simulated plans are at least as compliant with the VRA as the enacted plan. This should not be construed as a determination that the enacted plan is legally compliant with the VRA.
Indeed, some states, like Mississippi, demonstrate that minority opportunity districts can be drawn with smaller concentrations of Black voters.

As we provide the full code and data necessary to replicate the simulations, the sampling constraints can be easily altered to use alternative methods of VRA compliance, by changing a single line of code for each VRA constraint.
Indeed, this is just one approach to the VRA, which could alternatively include broader considerations, like primary elections, or more narrow considerations, like candidate-specific models of turnout.
Our approach is rooted in the effects side of VRA compliance based on the enacted plan, in that minority preferred candidates can win elections in as many minority opportunity districts as in the enacted plan.
It should not be construed as a determination of how many VRA districts a state should have drawn.

In simulating districts, there is often a trade-off between complying with constraints and obtaining a diverse sample of districts.

The SMC algorithm samples from the space of all possible redistricting plans which are contiguous and meet the population balance threshold of 0.5\%, and, depending on the state, an additional administrative unit splitting constraint.
Within this space, plans which are more compact and better align with other specified criteria have a higher probability of being drawn.
While the SMC algorithm has theoretical guarantees of sampling from the distribution specified by these constraints, in practice, it may become exceedingly difficult to draw plans which meet all the user-specified requirements and adhere to population parity and contiguity constraints.
In these cases, the algorithm may be stuck in a handful of plans, which get duplicated many times across the simulated set.
To ensure that the sampler is not getting stuck, or producing an unrepresentative sample, it is important to check several diagnostics.
We conduct extensive diagnostic checks reflecting the best practices.\cite{McCartanEAl:2021}

First, we check the final-stage resampling weights from the SMC algorithm to ensure that the constraints imposed on the sampling process are not too severe.
Next, we evaluate how \emph{diverse} the sample is by measuring the variation of information (VI) distance between plans.
If two plans place most people into the same districts, then the VI distance between them is low; if two plans assign people to districts very differently, the VI distance is larger.\cite{guth2020three}
A diverse sample should have a wide variety of redistricting plans, and therefore a high average pairwise VI distance.

Conversely, when the algorithm becomes stuck, many duplicated or highly-similar plans will show up as a very low pairwise VI distance.
After checking the weights and VI distance, we then use diagnostics from each iteration of the SMC procedure to ensure that there are no sampling bottlenecks or efficiency losses.
These bottlenecks can lead to a small number of valid districts in a certain area of a state to appear an abnormally large portion of the time in the final sample.
The \textit{redist} software has heuristic checks for these bottlenecks built in to its diagnostic reporting.
Finally, we evaluate the convergence of the algorithm for the specific set of summary statistics described above that are of interest to practitioners.
This is accomplished by splitting the sample across at least two independent runs of the sampling procedure; from this, we calculate the Gelman-Rubin \(\hat R\) statistic for each quantity of interest.\cite{gelman1992inference, vehtari2019rank}
The \(\hat R\) statistic will be large if the independent runs produce samples that are not alike.
We check that \(\hat R\) is around 1.05 or less for all of the calculated summary statistics, which indicates likely convergence.

After the simulations are completed, each diagnostic is presented in an automatically generated plot and summary report of the plan.
These are reviewed by a different member of the team through a public ``pull request'' on the repository.
Especially for larger states, many rounds of sampling and adjusting algorithm parameters such as the sample size and constraint strengths were often necessary to pass our battery of diagnostic checks.
Only once these checks were passed and validated by another member of the team was the sample admitted to the dataset. Figure \ref{fig:ga-validate} in the Appendix display these diagnostic checks for our Georgia simulations.

Our simulations therefore serve as a realistic set of alternative plans.
However, they do not represent our evaluation of the \emph{legality} of the enacted and other plans.
While simulations are used increasingly in litigation of districts, they are not always a deciding piece of evidence and simulations in cases often need to be tailored in specific ways that are difficult to characterize generally.
Our parameters of the simulation are open-source and replicable, and serve as a useful template for those interested in constructing their own simulation.

\hypertarget{usage-notes}{%
\section*{Usage Notes}\label{usage-notes}}
\addcontentsline{toc}{section}{Usage Notes}

The \fiftydata are well-suited to assess what types of partisan or racial outcomes could have happened under alternative plans in a given state.
Alternative plans should not be compared across states, for the same reason that comparing the metrics of enacted plans across states is also invalid.
A variant of a cross-state comparison could be made through computing within-state differences between enacted and simulated alternative plans, and comparing those differences across several states as a measure of gerrymandering.\cite{chen_cottrell_2016, kenny2022widespread}

As an example, we continue our exploration of Georgia's 2020 Congressional redistricting plan.
All of our data and analysis code is found in the open-source \texttt{R} package \texttt{alarmdata}\cite{alarmdata}.
After loading the package, we start by downloading a redistricting \texttt{redist\_map} object, which contains the basic information about the redistricting problem and Census and electoral variables as covariates.

\begin{Shaded}
\begin{Highlighting}[]
\FunctionTok{library}\NormalTok{(redist)}
\FunctionTok{library}\NormalTok{(alarmdata)}

\NormalTok{ga\_map }\OtherTok{\textless{}{-}} \FunctionTok{alarm\_50state\_map}\NormalTok{(}\StringTok{"GA"}\NormalTok{)}
\end{Highlighting}
\end{Shaded}

Separately, users can download the simulated plans with the \texttt{alarm\_50state\_plans()} function.
These plans come in the \texttt{redist\_plans} object format described in the Data Records Section, such that each row is a district in one draw of the simulation.
To pre-calculate common plan statistics for compactness, administrative boundary splits, and partisanship, we set the optional \texttt{stats\ =\ TRUE} argument.

\begin{Shaded}
\begin{Highlighting}[]
\NormalTok{plans }\OtherTok{\textless{}{-}} \FunctionTok{alarm\_50state\_plans}\NormalTok{(}\StringTok{"GA"}\NormalTok{, }\AttributeTok{stats =} \ConstantTok{TRUE}\NormalTok{)}
\end{Highlighting}
\end{Shaded}

Our redistricting objects are \texttt{tibble} objects, and therefore can be operated on by standard \texttt{tidyverse} functions \cite{tidyverse:2019}.
For example, the simulation plans object is also a \texttt{tibble} dataframe where one row is a district in a simulation.
Users can also create arbitrary statistics from the existing Census and VEST variables.
For example, below we calculate the Democratic proportion of the two-party vote share in the 2020 presidential election for each plan in Georgia.

\begin{Shaded}
\begin{Highlighting}[]
\NormalTok{plans }\OtherTok{\textless{}{-}}\NormalTok{ plans }\SpecialCharTok{\%\textgreater{}\%}
    \FunctionTok{mutate}\NormalTok{(}\AttributeTok{dem\_2020 =}\NormalTok{ pre\_20\_dem\_bid }\SpecialCharTok{/}\NormalTok{ (pre\_20\_dem\_bid }\SpecialCharTok{+}\NormalTok{ pre\_20\_rep\_tru))}
\end{Highlighting}
\end{Shaded}

Often, analysts will want to compare a specific enacted map or a counterproposal to our simulations.
The \fiftydata plan objects have the map enacted and finalized by each state as a reference plan in the draw \texttt{cd\_2020}, but analysts can also add custom reference plans.
Below, we add the 2010 Georgia Congressional plan to serve as a comparison for later analyses.
We pass our \texttt{plans} simulation object to \texttt{alarm\_add\_plan()}, and add the \texttt{cd\_2010} column as a reference redistricting plan.
The \texttt{name} argument provides a name for the reference plan in the plans object.

\begin{Shaded}
\begin{Highlighting}[]
\NormalTok{plans }\OtherTok{\textless{}{-}}\NormalTok{ plans }\SpecialCharTok{\%\textgreater{}\%} 
  \FunctionTok{alarm\_add\_plan}\NormalTok{(}\AttributeTok{map =}\NormalTok{ ga\_map, }\AttributeTok{ref\_plan =}\NormalTok{ ga\_map}\SpecialCharTok{$}\NormalTok{cd\_2010, }\AttributeTok{name =} \StringTok{"cd\_2010"}\NormalTok{)}
\end{Highlighting}
\end{Shaded}

Additionally, we present convenient visualization functions for immediate use with the \fiftydata.
For example, maps are easily created with the \texttt{redist::redist.plot.map()} function.
The \texttt{redist::redist.plot.hist()} function plots a variable in a plans object as a histogram while the \texttt{redist::redist.plot.distr\_qtys()} function generates a district-level summary statistic as a boxplot.\\

For example, to display the distribution of the two-party 2020 Democratic vote in each alternative district, we run:

\begin{Shaded}
\begin{Highlighting}[]
\FunctionTok{redist.plot.distr\_qtys}\NormalTok{(plans, }\AttributeTok{qty =}\NormalTok{ dem\_2020, }\AttributeTok{geom =} \StringTok{"boxplot"}\NormalTok{) }\SpecialCharTok{+}
    \FunctionTok{labs}\NormalTok{(}\AttributeTok{y =} \StringTok{"2020 Presidential Two{-}Party Vote Share"}\NormalTok{) }\SpecialCharTok{+}
    \FunctionTok{theme\_bw}\NormalTok{()}
\end{Highlighting}
\end{Shaded}

\begin{figure}

{\centering \includegraphics[width=0.8\linewidth]{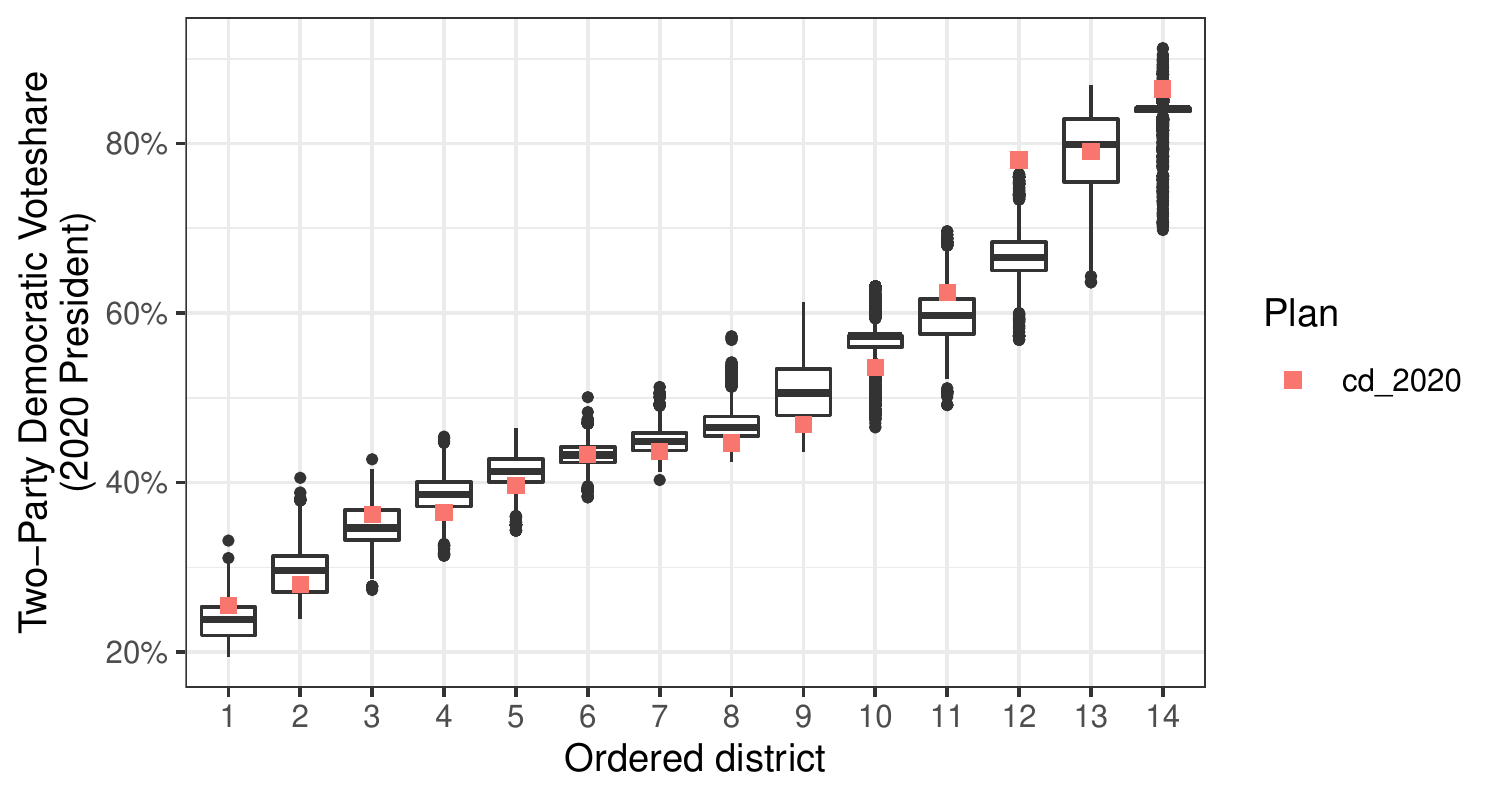} 

}

\caption{Boxplots of the two-party Democratic vote share in the 2020 presidential election in each of our 5,000 simulated congressional districts. Districts are ordered by their rank ordering of the Democratic vote within a plan, ranging from the least Democratic (left) to the most Democratic (right) district.  The solid bar shows the median Democratic vote, and black points indicate outlier values outside of the interquartile range. The red squares indicate the Democratic vote in the enacted plan. }\label{fig:ga-fig}
\end{figure}

Figure \ref{fig:ga-fig} shows the resulting boxplot, which indicates that, across most districts, the Democratic vote share in the enacted 2020 Congressional Plan falls within the range of our 5,000 simulated plans.
The two exceptions are the 12th and 14th ordered districts, both of which have Democratic vote shares larger than the expected range under our simulations.
This suggests that the enacted plan may be packing Democratic voters more than necessary by traditional criteria.
A similar analysis can be done with the proportion of the district Voting Age Population that is Black (BVAP), to detect signs of racial gerrymandering.
We hasten to note again that this result alone does not constitute conclusive evidence of partisan gerrymandering in the enacted plan.
The court may also require the simulated plans to conform to other guidelines, such as avoidance of pairing incumbents, which we did not implement here.
Nevertheless, the simulations serve as a diverse set of valid alternative maps that could be drawn to answer an otherwise intractable question: in what ways the enacted plan differs from other alternatives and by how much.

Further customization and visualization options are available in our companion packages \texttt{redist}\cite{redist}, which implements the simulation algorithm, and \texttt{redistmetrics}\cite{redistmetrics}, which efficiently computes metrics evaluating each plan.

\hypertarget{code-availability}{%
\section*{Code availability}\label{code-availability}}
\addcontentsline{toc}{section}{Code availability}

All code is publicly available on our GitHub repository for the \fiftydata project, \url{https://github.com/alarm-redist/fifty-states}.
The package \texttt{alarmdata}\cite{alarmdata} is a user-facing R package to download and work with our plans on the repository and our Dataverse.

\hypertarget{table-of-state-rules}{%
\section{Table of State Rules}\label{table-of-state-rules}}

Redistricting rules vary across states.
However, all states require districts to be contiguous and have approximately equal populations.
Further, many states include references to avoid splitting particular geographic areas, such as counties or ``communities of interest.''
As a result, we use a set of common ``base constraints'' applied to every state, and incorporate additional constraints for particular state circumstances (e.g., Colorado requires competitive districts, Louisiana requires preserving the cores from previous districts, etc.).

Our base constraints are contiguity, approximately equal populations, and a county / municipality constraint to avoid splitting.
Further, districts are drawn to be compact and are compared with the metrics described in the main text, unless the state explicitly mentions a measure that should be used (e.g., Iowa).
Consistent with laws in many states, none of our districts are drawn to favor incumbents or a political party.

Some states, such as Virginia and Pennsylvania, have residential segregation by race which naturally ensures that minority opportunity districts are included from a race-blind simulation process. While we check performance as a validation step, no additional constraints are needed within the simulations to ensure that compliance.

\begin{figure}
\includegraphics[width=1\linewidth]{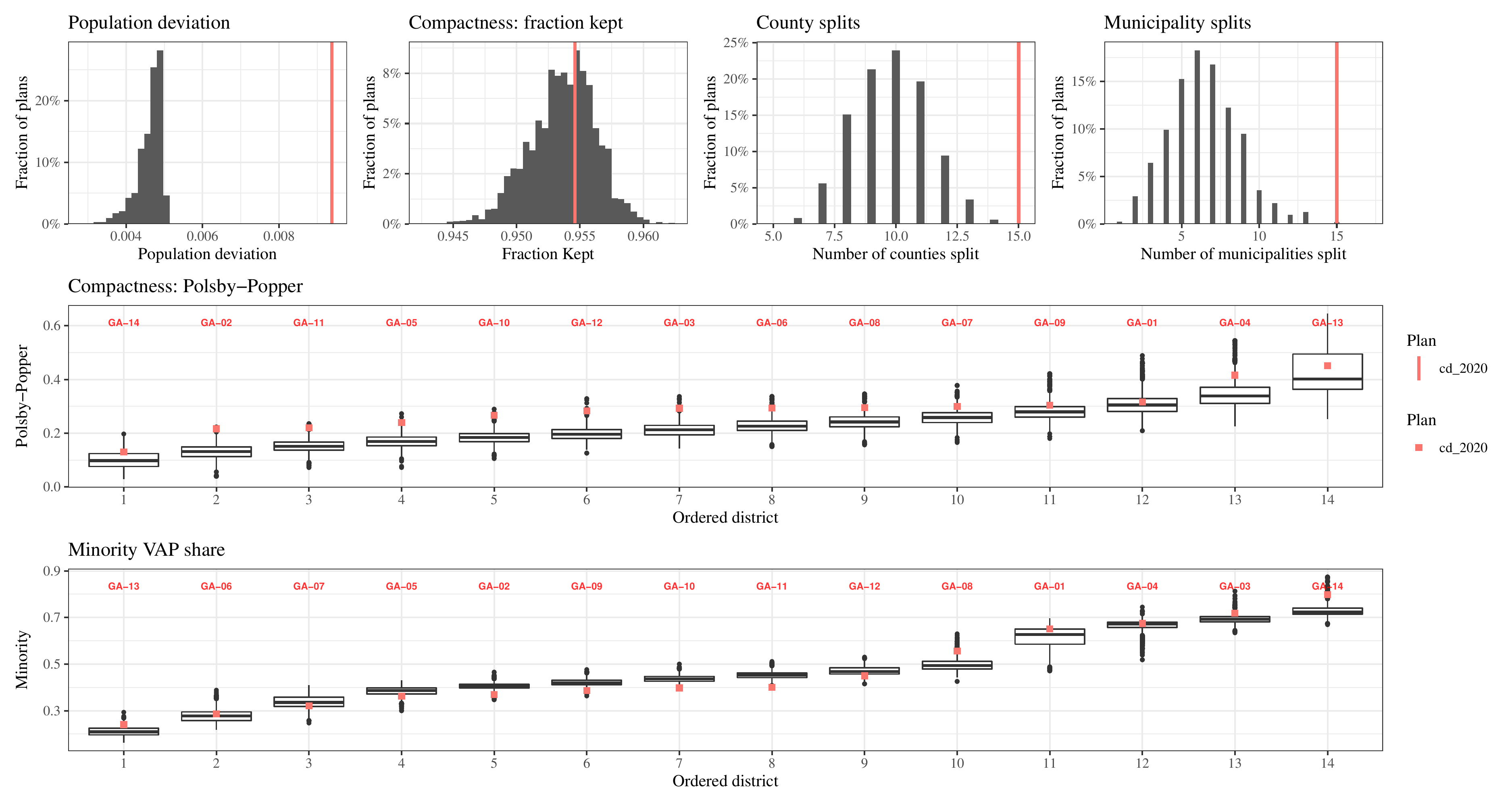} \caption{An example validation plot from our Georgia analysis. Analogous validation plots were made for all states, and include information on metrics discussed above like compactness, diversity, boundary splits, and minority VAP. Validation plots for all analyses are available in the individual state pull requests on our public repository.}\label{fig:ga-validate}
\end{figure}

\bibliography{references.bib}

\hypertarget{acknowledgements}{%
\section*{Acknowledgements}\label{acknowledgements}}
\addcontentsline{toc}{section}{Acknowledgements}

The authors thank the Harvard Data Science Initiative and Microsoft for computational support.

\hypertarget{author-contributions-statement}{%
\section*{Author contributions statement}\label{author-contributions-statement}}
\addcontentsline{toc}{section}{Author contributions statement}

C.T.K., C.M., T.S. and K.I. conceived the project, C.T.K., C.M., T.S., G.G., K.W., M.W., and S.K. conducted the data analysis. All authors contributed to the writing.

\hypertarget{competing-interests}{%
\section*{Competing Interests}\label{competing-interests}}
\addcontentsline{toc}{section}{Competing Interests}

C.T.K. has served as a paid expert for the Maryland Redistricting Commission. K.I. has served as a paid expert witness in the court cases related to the 2020 Congressional redistricting in Alabama, Kentucky, Ohio, and South Carolina.

\end{document}